\documentclass[
 aip,
 amsmath,amssymb,
 reprint,%
]{revtex4-1}

\usepackage{graphicx}
\usepackage{dcolumn}
\usepackage{bm}

\usepackage[utf8]{inputenc}
\usepackage[T1]{fontenc}
\usepackage{mathptmx}
\usepackage{etoolbox}
\usepackage{upgreek}
\usepackage{multirow}
\usepackage{natbib}
\usepackage[utf8]{inputenc}
\usepackage{color}
\usepackage{stackengine}
\usepackage[normalem]{ulem}
\UseRawInputEncoding

\makeatletter
\def\@email#1#2{%
 \endgroup
 \patchcmd{\titleblock@produce}
  {\frontmatter@RRAPformat}
  {\frontmatter@RRAPformat{\produce@RRAP{*#1\href{mailto:#2}{#2}}}\frontmatter@RRAPformat}
  {}{}
}%
\makeatother

\begin{document}

\preprint{AIP/123-QED}

\title{Impact of AlN buffer thickness on electrical and thermal characteristics of AlGaN/GaN/AlN HEMTs}

\author{Minho Kim}
\affiliation{Center for III-Nitride Technology, C3NiT - Janz\'en and Department of Physics, Chemistry and Biology (IFM), Link\"{o}ping University, SE-58183 Link\"{o}ping, Sweden}
\affiliation{Wallenberg Initiative Materials Science for Sustainability (WISE), Department of Physics, Chemistry and Biology (IFM), Link\"{o}ping University, SE-58183 Link\"{o}ping, Sweden}

\author{Dat Q. Tran}
\affiliation{Center for III-Nitride Technology, C3NiT - Janz\'en and Department of Physics, Chemistry and Biology (IFM), Link\"{o}ping University, SE-58183 Link\"{o}ping, Sweden}
\affiliation{Department of Electrical Engineering, Stanford University, Stanford, CA 94305, USA}

\author{Plamen P. Paskov}
\affiliation{Center for III-Nitride Technology, C3NiT - Janz\'en and Department of Physics, Chemistry and Biology (IFM), Link\"{o}ping University, SE-58183 Link\"{o}ping, Sweden}

\author{U.Choi}
\affiliation{Power and Wide-Band-Gap Electronics Research Laboratory, Institute of Electrical and Micro Engineering, \'{E}cole Polytechnique F\'ed\'erale de Lausanne, 1015 Lausanne, Switzerland}

\author{O.Nam}
\affiliation{Convergence Center for Advanced Nano Semiconductor (CANS), Department of Nano-Semiconductor, Tech University of Korea, 15073 Siheung-si, Korea} 

\author{Vanya Darakchieva}
\affiliation{Center for III-Nitride Technology, C3NiT - Janz\'en and Department of Physics, Chemistry and Biology (IFM), Link\"{o}ping University, SE-58183 Link\"{o}ping, Sweden}
\affiliation{Wallenberg Initiative Materials Science for Sustainability (WISE),Department of Physics, Chemistry and Biology (IFM), Link\"{o}ping University, SE-58183 Link\"{o}ping, Sweden} 
\affiliation{Terahertz Materials Analysis Center and  NanoLund and Solid State Physics, Lund University, 22100 Lund, Sweden}
\email{Authors to whom correspondence should be addressed: minho.kim@liu.se, vanya.darakchieva@liu.se}

\date{\today}

\begin{abstract}
We investigate the influence of AlN buffer thickness on the structural, electrical, and thermal properties of AlGaN/GaN high-electron mobility transistors (HEMTs) grown on semi-insulating SiC substrates by metal-organic chemical vapor deposition. X-ray diffraction and atomic force microscopy reveal that while thin AlN layers (120~nm) exhibit compressive strain and smooth step-flow surfaces, thicker single-layer buffers (550~nm) develop tensile strain and increased surface roughness. Multi-layer buffer structures up to 2~$\mu$m alleviate strain and maintain surface integrity. Low-temperature Hall measurements confirm that electron mobility decreases with increasing interface roughness, with the highest mobility observed in the structure with a thin AlN buffer. Transient thermoreflectance measurements show that thermal conductivity (ThC) of the AlN buffer increases with the thickness, reaching 188~W/m.K at 300~K for the 2~$\mu$m buffer layer, which is approximately 60\% of the bulk AlN ThC value. These results highlight the importance of optimizing AlN buffer design to balance strain relaxation, thermal management, and carrier transport for high-performance GaN-based HEMTs.
\end{abstract}

\maketitle
Gallium nitride (GaN)-based high-electron-mobility transistors (HEMTs) are among the most widely used semiconductor devices, with growing demand for improved performance to meet the needs of advanced radio frequency (RF) and power-switching applications.\cite{Tsao_2018}
AlGaN/GaN/AlN double-heterostructures (DH), incorporating AlN as a buffer layer, present strong potential for such applications due to AlN's wide bandgap (6.1~eV), high thermal conductivity (340~W/m.K), and enhanced two-dimensional electron gas (2DEG) confinement.\cite{Kim_2023,Kotani_2023,Wolf_2024} 
In addition, AlN buffers support high breakdown voltages and improve dynamic and noise performance by mitigating GaN buffer-related traps typically associated with intentionally introduced carbon or iron.\cite{Im_2022,Im_2024}
Recently, AlN-buffered HEMT structures have been actively studied on various substrates, including sapphire,\cite{Choi_2020a,Kim_2021}, SiC,\cite{Kim_2023,Li_2014}, and AlN substrates.\cite{Qi_2017, Kim_2025,Chen_2025} 
Among these, epitaxial growth on SiC remains challenging due to the thermal expansion mismatch with AlN, which limits the growth of high-quality thick films. However, our previous work has demonstrated that AlN-on-SiC growth is feasible through optimized growth techniques.\cite{Kim_2022}
In these structures, the buffer layer plays a critical role in promoting strain relaxation, reducing defect density, and facilitating thermal dissipation. These factors are closely related to the crystalline quality of the GaN channel, which strongly influences the 2DEG mobility, a key parameter in determining overall device performance. Furthermore, the thermal conductivity of the buffer layer significantly affects heat dissipation during high power operation, further emphasizing its importance for reliable and efficient device operation.\cite{Elwaradi_2023,Choi_2020a}
Among the key parameters that govern the device performance, the thickness of the AlN buffer layer plays a particularly important role. The thickness of AlN buffer directly affects strain relaxation, dislocation density, and thermal transport in the heterostructure. Thin AlN layers often fail to sufficiently relieve strain, leading to high defect densities and low thermal conductivity due to the strong phonon boundary scattering.\cite{Belkerk_2014,Tran_2020_APL} On the other hand, excessively thick buffers can result in growth-related problems such as crack formation. Despite its significance, the comprehensive impact of AlN buffer thickness on structural and thermal behavior has not been systematically studied.

In this study, we investigate the influence of AlN buffer thickness on the structural and thermal properties of AlGaN/GaN HEMTs grown on semi-insulating SiC substrates. X-ray diffraction (XRD), atomic force microscopy (AFM), and low-temperature Hall measurements are used to examine how variations in buffer thickness impact strain relaxation, defect density, and 2DEG characteristics. Additionally, thermal conductivity of the HEMT structures is evaluated using transient thermoreflectance (TTR).

\begin{figure}[htbp]
\includegraphics[keepaspectratio=true,width=1\linewidth, clip, trim=0.1cm 0.5cm 0.1cm 0.1cm,scale=0.8 ]{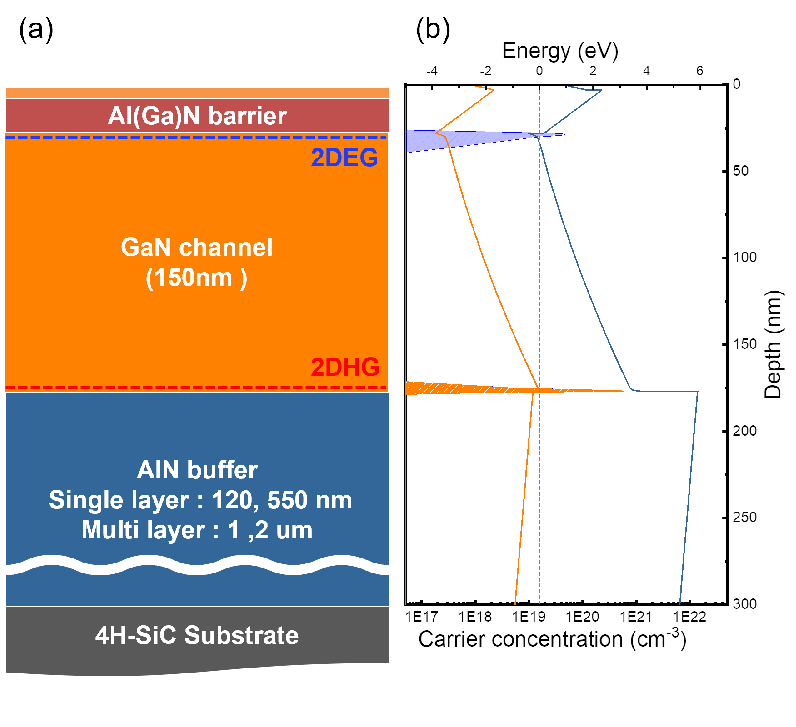}
\caption{(a) Schematic cross-sectional structure of the AlGaN/GaN/AlN HEMT on a 4H-SiC substrate. The structure includes an AlGaN barrier, a 150 nm GaN channel layer, and an AlN buffer layer with various thicknesses (120, 550, 1000 and 2000 nm). (b) Energy-band diagram indicating the formation of a 2DEG at the AlGaN/GaN interface and a 2DHG at the GaN/AlN interface.}
\label{fig:schematic}
\end{figure}

All samples were grown on semi-insulating SiC (SI-SiC) substrates by metal-organic chemical vapor deposition (MOCVD), using trimethylgallium (TMGa) and trimethylaluminum (TMAl) as Ga and Al precursors, respectively, with H$_2$ and N$_2$ as carrier gases. After high-temperature hydrogen cleaning of the SiC surface, epitaxial growth was carried out at 1250~$^\circ$C. The AlN buffer layer was implemented either as a single layer (120~nm and 550~nm) or as a multi-layer structure (1~$\mu$m and 2~$\mu$m). To mitigate cracking due to thermal expansion mismatch between AlN and the SiC substrate in multi-layer structures, a three-dimensional buffer design was employed. Detailed growth conditions are described in our previous work.\cite{Kim_2022} The epitaxial stack consists of a 2~nm GaN cap layer, a 25 nm AlGaN barrier, a 150 nm GaN channel, an AlN insert layer, and an AlN buffer layer, as illustrated in Fig.~\ref{fig:schematic}(a).

\begin{figure*}[htbp]
 \centering
\includegraphics[keepaspectratio=true,width=0.9\linewidth, clip, trim=0.1cm 1.3cm 0.1cm 0.3cm]{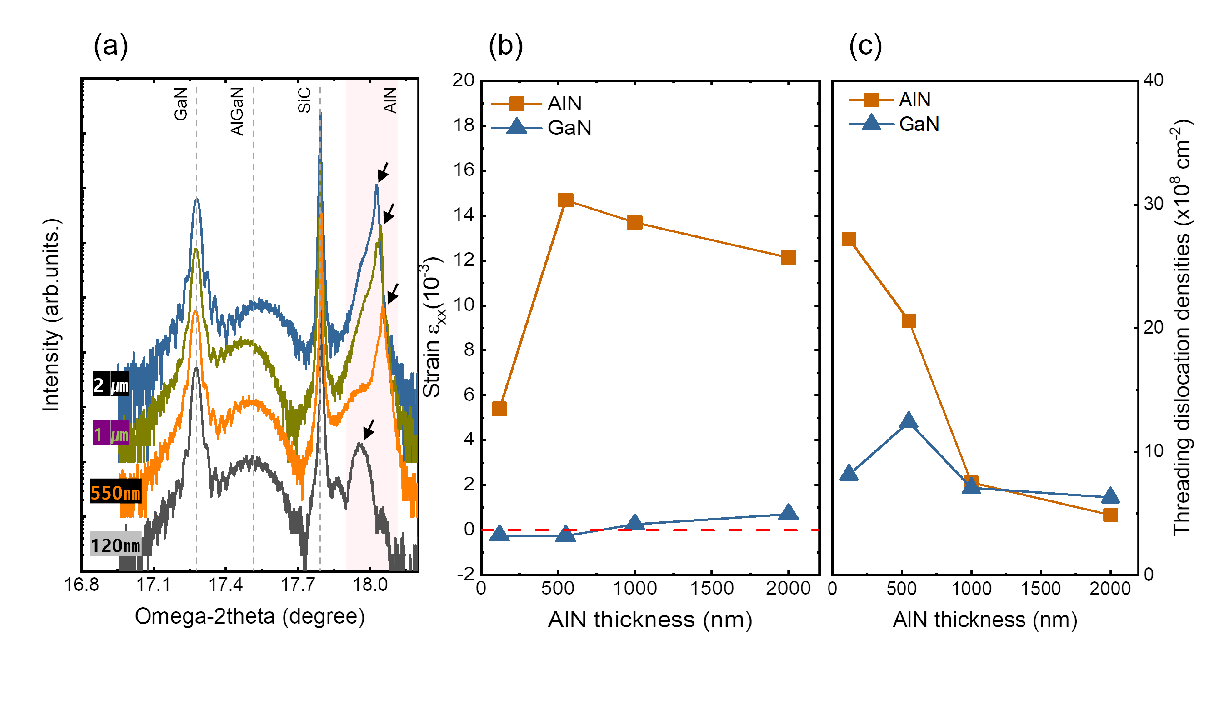}
\caption{(a) XRD $\omega$-2$\theta$ scans for the HEMTs samples with varying AlN buffer layer thicknesses. (b) In-plane strain ($\varepsilon_{xx}$) of AlN and GaN layers as a function of AlN buffer thickness. (c) The threading dislocation density in the AlN buffer and GaN channel layers.}  
\label{fig:2theta}
\end{figure*}

For structural analysis, XRD $\omega$-2$\theta$ scans were performed. As shown in Fig.~\ref{fig:2theta}(a), the diffraction peaks corresponding to GaN, AlGaN, SiC, and AlN are clearly observed for all samples with varying AlN buffer thicknesses. The AlN peak becomes more intense and exhibits a narrower full width at half maximum (FWHM) with increasing thickness, indicating improved crystallinity of the AlN layer. In addition, for the 550~nm and thick multi-layer samples, two overlapping AlN (0002) peaks are observed. This indicates the coexistence of domains with different strain states-specifically, a partially relaxed upper layer and a lower layer that remains under compressive stress\cite{Darakchieva_2004} due to the lattice and thermal mismatch with the SiC substrate.\cite{Zollner_2019} The effects of strain are more clearly seen in Fig.~2(b), which displays the shift of the (002) diffraction peak for AlN and GaN relative to their relaxed positions. A narrow, dominant peak is observed for AlN (indicated by the black arrow), which was used to evaluate the strain in the thicker AlN layers. The $c$-axis lattice constant was obtained from the symmetric (002) $\omega$-$2\theta$ scan by applying Bragg's law to the (002) reflection. The $c$-axis lattice parameters used for reference were $c_{o}=5.1852$~\AA\ for GaN and $c_{o}=5.01$~\AA\ for AlN.\cite{Darakchieva_2007,Xie_2012}

The out-of-plane strain component was then calculated as

\begin{equation}
\varepsilon_{zz}=\frac{c-c_{o}}{c_{o}}
\end{equation}
where $c_{o}$ denotes the relaxed $c$-axis lattice constant of the film material. The in-plane strain component was obtained from $\varepsilon_{zz}$ via the elastic constants according to

\begin{equation}
\varepsilon_{xx}=-\frac{C_{33}}{2C_{13}}\,\varepsilon_{zz}
\end{equation}
We used the following stiffness constants (in GPa): for AlN, $C_{33}=356$ and $C_{13}=98$; for GaN, $C_{33}=381$ and $C_{13}=114$.\cite{Yamaguchi_1997,Xie_2012}

As the AlN thickness increases from 120 nm to 500 nm, a clear transition is observed. The 120 nm-thick AlN layer, grown directly on the SiC substrate, exhibits a strong compressive strain, while the 550~nm sample, beyond which cracking occurs, shows a tensile strain, mainly due to the thermal expansion mismatch between AlN and the SiC substrate.\cite{Kim_2022} Under current growth conditions, 550~nm appears to be the practical thickness limit for a single AlN layer without inducing cracks. In comparison, the 1~$\mu$m and 2~$\mu$m samples, fabricated using a multilayer buffer approach, exhibit a reduced overall strain compared to the 550~nm single-layer, indicating that thicker AlN films can be grown more reliably through strain-relief engineering. In all samples, the in-plane strain in the GaN channel is near zero, indicating that although the GaN is nominally subject to compressive misfit with the underlying AlN, it is largely relaxed owing to the ~150-nm channel thickness. This observation is further supported by the fact that while the crystalline quality of the AlN buffer improves with increasing thickness, the GaN channel quality remains relatively constant (Fig.~\ref{fig:2theta}(c)). This highlights the need for further investigation into how the structural properties of the GaN channel, whether strained or relaxed, affect its electron transport properties.

\begin{figure}[htbp]
\includegraphics[keepaspectratio=true,width=0.9\linewidth, clip, trim=0.0cm 0.0cm 0.1cm 0.1cm,scale=0.7 ]{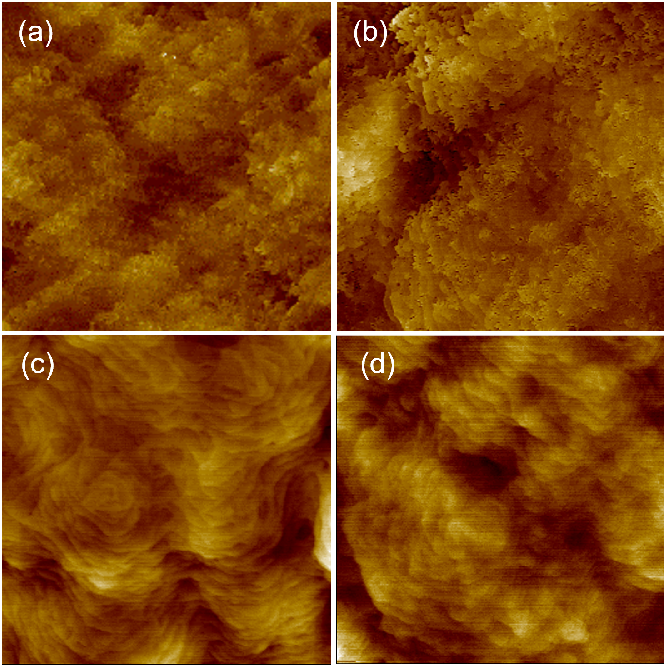}
\caption{Atomic force microscopy (AFM) surface images of GaN layers grown on AlN buffer layers with different thicknesses: (a) 120 nm, (b) 550 nm, (c) 1000 nm, and (d) 2000 nm. All scans are performed over a $5 \times 5~\mu\mathrm{m}^2$ area.}
\label{figure:afm}
\end{figure}

\begin{figure*}[htbp]
\includegraphics[keepaspectratio=true,width=0.9\linewidth, clip, trim=0.1cm 1.0cm 0.1cm 0.5cm ]{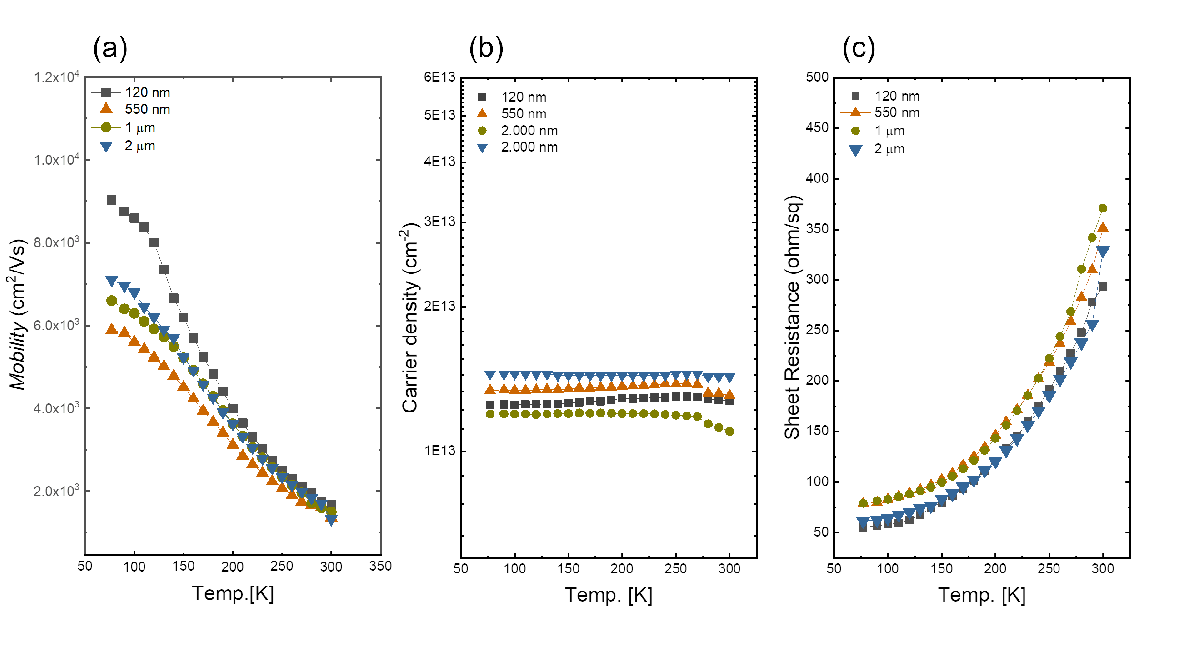}
\caption{Temperature-dependent Hall measurements of GaN layers grown on AlN buffer layers with different thicknesses (120 nm, 550 nm, and 2000 nm): (a) Electron mobility, (b) Sheet carrier density, and (c) Sheet resistance. The measurements were performed from 77 K to 300 K. The sample with a 2000 nm AlN buffer shows the highest mobility and lowest sheet resistance at low temperatures.}
\label{fig:hall}\cite{}
\end{figure*}

Figure~\ref{figure:afm} shows AFM surface images of the HEMT structures. The sample with a thin 120 nm AlN buffer exhibits an angled step-flow morphology, likely resulting from the conformal growth of the AlN layer along the substrate atomic steps. In contrast, this step structure is not observed in the rest of the samples with thicker AlN buffer layers (550~nm, 1~$\mu$m, and 2~$\mu$m). This is consistent with the tensile stress measured in the respective AlN layers (Fig.~\ref{fig:2theta}(b)). In multilayer AlN structures, the initial low-temperature AlN islands merge into larger grains during growth as deduced from temperature-dependent growth experiments.~\cite{Kim_2022} Nonetheless, these thicker multilayer samples exhibit slightly improved root mean square (RMS) surface roughness.

Figure~\ref{fig:hall} presents the temperature-dependent Hall characteristics measured from 77\,K to 300\,K. From left to right, the plots show variations in 2DEG mobility and density, and sheet resistance. The HEMT with a thin, single-layer 120~nm AlN buffer exhibits the highest mobility and overall electrical performance. In contrast, the thick single-layer buffer (550 nm) structure shows the lowest mobility, likely due to the increased dislocation density in the GaN channel and the higher roughness of the interface. The two-layer buffer structure demonstrates slightly improved electrical properties compared to the thick single-layer case. This trend correlates well with the AFM surface morphology shown in Fig.~\ref{figure:afm}, especially in terms of mobility behavior. Since 2DEG mobility is highly sensitive to interface roughness, the smoother surface observed in the thin AlN sample likely contributes to its superior electrical characteristics.

The summarized structural and electrical properties of the investigated HEMT structures are provided in Table~\ref{table}. Thin AlN layers offer advantages in terms of growth simplicity and cost-effectiveness. Similar structures have been reported in previous studies,~\cite{Chen_2018, Chen_2023} showing performance comparable to conventional GaN buffer-based HEMTs. However, limitations remain in enhancing breakdown voltage and suppressing carrier trapping, which are critical for ensuring long-term device reliability. These issues are particularly relevant when considering the thermal characteristics discussed below.

\begin{table*}[htbp]
\centering
\caption{Structural, electrical, and surface properties of the HEMT structures as a function of AlN buffer layer thickness}
\label{table}
\begin{tabular}{c |cc|cc|cc|c |ccc|ccc}
\toprule
\textbf{AlN} & \multicolumn{2}{c|}{\textbf{FWHMs (002)/(102)}} & \multicolumn{2}{c|}{\textbf{TDD}} & \multicolumn{2}{c|}{$\boldsymbol{\varepsilon_{xx}}$} & \textbf{RMS} & \multicolumn{3}{c|}{\textbf{300 K}} & \multicolumn{3}{c}{\textbf{77 K}} \\
\textbf{t (nm)} & AlN & GaN & AlN & GaN & AlN & GaN & (nm) & Mobility & $N_s$ & $R_s$ & Mobility & $N_s$ & $R_s$ \\
& \multicolumn{2}{c|}{(arcsec)} & \multicolumn{2}{c|}{($\times 10^8$ cm$^{-2}$)} & \multicolumn{2}{c|}{($\times 10^{-3}$)} & & (cm$^2$/Vs) & ($\times 10^{13}$ cm$^{-2}$) & ($\Omega/\square$) & (cm$^2$/Vs) & ($\times 10^{13}$ cm$^{-2}$) & ($\Omega/\square$) \\
\hline
120   & 580/935 & 445/490 & 27.2 & 8.14 & -4.9  & -2.7 & 0.35  & 1670 & 1.27 & 294 & 9030 & 1.24 & 55.3 \\
550   & 270/860 & 350/660 & 20.6 & 12.4 & 4.1  & -2.7 & 0.63  & 1357 & 1.31 & 351 & 5890 & 1.34 & 79.1 \\
1000 & 190/500 & 165/530 & 7.12 & 7.5  & 2.5   & -1.7 & 0.40 & 1500 & 1.12 & 371 & 6600 & 1.20 & 79.1 \\
2000 & 210/405 & 280/465 & 4.88 & 6.31 & 1.6  & -2.1 & 0.40  & 1326 & 1.43 & 329 & 7100 & 1.44 & 61.5 \\
\hline
\hline
\end{tabular}
\end{table*}

\begin{figure}[htbp]
\includegraphics[keepaspectratio=true,width=1.1\linewidth, clip, trim=1.5cm 1.0cm 0.1cm 1.0cm ]{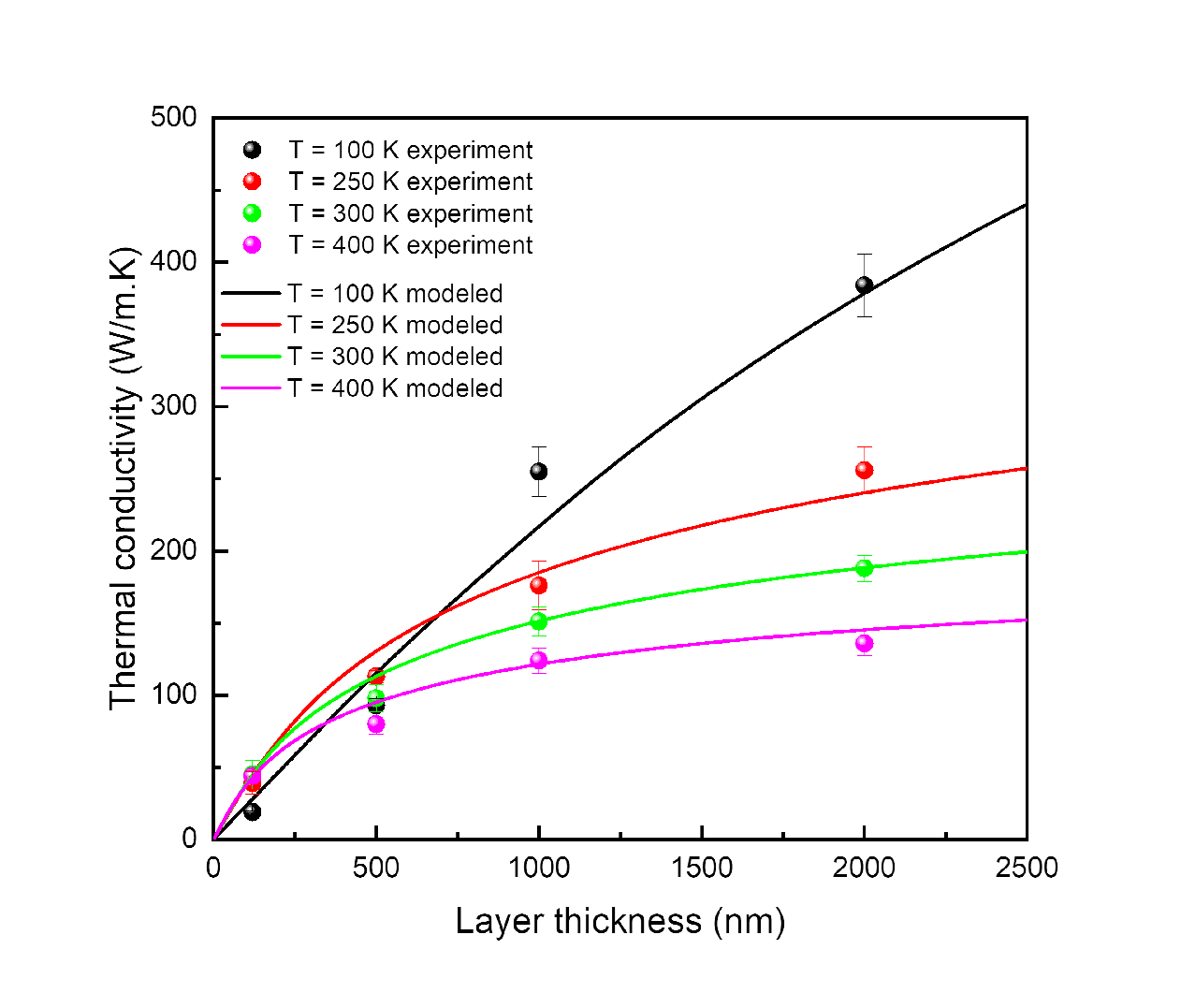}
\caption{Thermal conductivity of the AlN buffer layers with thickness of 120 nm, 550 nm, 1 $\mu$m and 2 $\mu$m, measured in the temperature range of 100 - 400 K. The solid lines present results from calculations based on the modified Callaway model.}
\label{fig:thermal}
\end{figure}

The thermal performance of GaN HEMTs is a critical factor, particularly in high-power and high-frequency applications. Therefore, the thermal conductivity ($k$) of the device components must be considered in the design process to ensure efficient heat dissipation and prevent thermal degradation. The thermal conductivity measurements are carried out by the pump-probe TTR method, which has been described in our previous works.\cite{Tran_2020_APL, Tran_2022_PRM} Since in the HEMT structures studied here only the AlN buffer is varying, we concentrate on the $k$ of this layer. Figure~\ref{fig:thermal} shows the thermal conductivity of AlN layers as a function of thickness at different temperatures. It is found that thinner AlN layers exhibit a lower $k$ regardless of the temperature. The thermal conductivity decreases with thickness and is more pronounced at lower temperatures. This behavior is attributed to the phonon-boundary scattering. At low temperatures, the average phonon mean-free-path becomes longer and the phonons experience stronger scattering at the interfaces, significantly reducing $k$.\cite{Tran_2020_APL} In thicker layers, the effect of phonon boundary scattering diminishes and three-phonon (Umklapp) scattering becomes the dominant mechanism that governs $k$.\cite{Tran_2022_PRM} This results in a notable increase in $k$ with thickness. For a 2 $\mu$m-thick AlN layer, $k$ = 190 W/m.K was obtained at 300 K, which corresponds to approximately 60\% of the bulk value for AlN (321 W/m.K).\cite{Cheng_2020} As it is seen from Fig. \ref{fig:thermal}, the experimental data are in good agreement with the calculations based on the modified Callaway model,\cite{Tran_2022_PRM} where the Umklapp, dislocation and boundary phonon scattering are taken into account.

The use of thicker AlN buffer layers (with higher $k$) in HEMT structures is expected to be beneficial for heat dissipation because it will ($i$) lower the maximum temperature at the top surface and ($ii$) minimize the effect of thermal boundary resistance at the AlN/GaN interface located far away from the hot spot at the top of the structure.\cite{Tran_2025}

In summary, we examined the influence of AlN buffer thickness on the structural, electrical, and thermal properties of AlGaN/GaN HEMTs grown on semi-insulating SiC substrates. HEMTs with thin AlN buffers exhibited smooth surfaces and high electron mobility, whereas those with thicker AlN layers showed improved crystallinity but exhibited slightly increased interface roughness and reduced mobility. Multi-layer buffer structures effectively mitigated strain in thicker AlN layers while maintaining reasonable surface quality. The thermal conductivity of the AlN buffer layer increased with the thickness, implying an enhanced thermal transport across the HEMT structures with thicker AlN buffer. These findings highlight the importance of optimizing buffer layer design to achieve a balance between electrical performance and thermal management in high-power GaN-based devices.

\begin{acknowledgments}
This work is supported by Knut and Alice Wallenberg Foundation funded grant 'Transforming ceramics into next generation semiconductors' (Grant No. 2024.0121), by the Swedish Governmental Agency for Innovation Systems VINNOVA under the Competence Center Program (Grant No. 2022-03139), by the Wallenberg Initiative Materials Science for Sustainability (WISE) funded by the Knut and Alice Wallenberg Foundation, by the Swedish Research Council (VR) under Grant No.2023-04993, and by the Swedish Government Strategic Research Area NanoLund and in Materials Science on Functional Materials at Link\"oping University, Faculty Grant SFO Mat LiU No. 009-00971. V.D. acknowledges support by the Knut and Alice Wallenberg Foundation for a Scholar award (Grant No. 2023.0349) and D.Q.T acknowledges support by the Knut and Alice Wallenberg Foundation for a Postdoctoral Fellowship to Stanford University.
\end{acknowledgments}

\section*{Data Availability Statement}
The data that support the findings of this study are available from the corresponding author upon reasonable request.

\section*{references}
\bibliography{aipsamp}

\end{document}